\documentclass[twocolumn,a4paper,prc,floatfix,showpacs,preprintnumbers,
	nofootinbib]{revtex4}
\usepackage[english]{babel}
\usepackage{mathrsfs}
\usepackage{amssymb}
\usepackage{amsmath}
\usepackage[dvips]{graphicx}
\begin{document}

\author{T. Lappi}

\email[Email address: ]{tuomas.lappi@helsinki.fi}
\affiliation{
Department of Physical Sciences,
Theoretical Physics Division
}
\affiliation{
Helsinki Institute of Physics\\
P.O. Box 64,
FIN-00014 University of Helsinki,
Finland
}

\title{Rapidity distribution of gluons in the classical
field model for  heavy  ion collisions}


\preprint{HIP-2004-49/TH}
\preprint{hep-ph/0409328}

\newcommand{\roots}{\sqrt{s}}

\newcommand{\xt}{\mathbf{x}_T}
\newcommand{\yt}{\mathbf{y}_T}
\newcommand{\pt}{{\mathbf{p}_T}}
\newcommand{\ptt}{p_T} % scalar
\newcommand{\qt}{{\mathbf{q}_T}}
\newcommand{\kt}{{\mathbf{k}_T}}
\newcommand{\lt}{{\mathbf{l}_T}}
\newcommand{\Dt}{{\mathbf{D}_T}}
\newcommand{\At}{{\mathbf{A}_T}}
\newcommand{\nabt}{\boldsymbol{\nabla}_T}

\newcommand{\ptil}{\tilde{p}}
\newcommand{\ktil}{\tilde{k}}
\newcommand{\qtil}{\tilde{q}}

\newcommand{\emu}{\! e_\mu \!}
\newcommand{\enu}{\! e_\nu \!}

\newcommand{\ud}{\, \mathrm{d}}
\newcommand{\uc}{{\mathrm{c}}}
\newcommand{\ul}{{\mathrm{L}}}
\newcommand{\intd}{\int \!}
\newcommand{\tr}{\, \mathrm{Tr} \, }
\newcommand{\R}{\mathrm{Re}}
\newcommand{\nc}{{N_\mathrm{c}}}
\newcommand{\nf}{{N_\mathrm{F}}}
\newcommand{\half}{\frac{1}{2}}
\newcommand{\hc}{\mathrm{\ h.c.\ }}
\newcommand{\nosum}[1]{\textrm{ (no sum over } #1 )}
\newcommand{\na}{\, :\!}
\newcommand{\nb}{\!: \,}
\newcommand{\cf}{C_\mathrm{F}}
\newcommand{\ca}{C_\mathrm{A}}
\newcommand{\df}{d_\mathrm{F}}
\newcommand{\da}{d_\mathrm{A}}
\newcommand{\nr}[1]{(\ref{#1})} 
\newcommand{\dadj}{D_{\mathrm{adj}}}
\newcommand{\ra}{R_A}

\newcommand{\re}{\textcolor{red}}
\newcommand{\gr}{\textcolor{green}}
\newcommand{\bl}{\textcolor{blue}}    

\newcommand{\gev}{\ \textrm{GeV}}
\newcommand{\fm}{\ \textrm{fm}}
\newcommand{\ls}{\Lambda_\mathrm{s}}
\newcommand{\qs}{Q_\mathrm{s}}
\newcommand{\lqcd}{\Lambda_{\mathrm{QCD}}}
\newcommand{\as}{\alpha_{\mathrm{s}}}

\newcommand{\init}{\textrm{init}}
\newcommand{\final}{\textrm{final}}

%%%%%%%%% commands from KK/M. Laine
\newcommand{\fig}{Fig.~}
\newcommand{\eq}{Eq.~}
\newcommand{\se}{Sec.~}
\newcommand{\eqs}{Eqs.~}

\begin{abstract}
The rapidity distribution of gluons produced in heavy ion collisions
is studied by a numerical computation in 2+1-dimensional classical
Yang-Mills theory. By assuming that
the classical source strength $g^2 \mu$ depends on rapidity as
$g^4 \mu^2 \sim e^{\pm \lambda y}$ and studying collisions of
two nuclei with different $g^2 \mu$ we find that the rapidity
distribution of produced gluons at central rapidities is very broad.
The transverse energy is seen to decrease even more slowly
as a function of $y$ than the multiplicity.
We discuss these results and the range in $y$ and $\roots$
where they are applicable in the light of experimental results
and other theoretical calculations.
\end{abstract}

\maketitle

\section{Introduction}

The longitudinal distribution of particles produced in a relativistic heavy
ion collision is experimentally most easily measured as a function of
the pseudorapidity $\eta$, because measuring pseudorapidity does not require
particle identification. This measurement has been done for a wide range
in $\eta$ e.g. by the PHOBOS experiment \cite{Back:2001bq}.
Theoretical calculations, on the other hand, naturally produce
rapidity distributions. Given the different particle species and the
different stages of the collision from the initial conditions to
hadronisation and decoupling, finding the right way to transform between
pseudorapidity and rapidity is not always simple.
The uncertainty in interpreting
experimental pseudorapidity distributions in terms of
theoretical calculations of rapidity dependence is discussed in e.g.
\cite{Tuominen:2002sq}.
More recently, however, the BRAHMS \cite{Bearden:2004yx} and STAR
\cite{Adams:2003xp} experiments have
measured charged particle yields also as a function of
the rapidity $y$. Especially the BRAHMS data, covering a wide range in $y$,
facilitates the comparison between theoretical models and experimental
results.

The aim of this paper is to calculate numerically the rapidity dependence
of gluon production in the classical field model around central rapidities.
In \se\ref{sec:model} we briefly review this model
and how it can be applied to study gluon production in heavy ion collisions.
In  \se\ref{sec:sat} we discuss the relation between the McLerran-Venugopalan
model and saturation and argue that the rapidity dependence
of particle production can be studied by varying the strengths of
the classical sources as $g^4 \mu^2 \sim e^{\pm \lambda y}$,
at least for large enough $\roots$ and for small enough rapidity.
Numerical results from applying the classical field model to
collisions of two nuclei with different color charge densities
are presented in \se\ref{sec:results} and
these results and their applicability to heavy ion phenomenology
discussed in \se\ref{sec:pheno}.

\section{The classical field model}\label{sec:model}

The McLerran-Venugopalan model for the small $x$ wavefunction of an
ultrarelativistic nucleus was suggested in
\cite{McLerran:1994ni,McLerran:1994ka,McLerran:1994vd}.
The classical field model for the initial stage of a collision of two heavy ions,
based on the McLerran-Venugopalan model for the nuclear wavefunction,
was formulated in \cite{Kovner:1995ts} and in
\cite{Kovchegov:1997ke,Kovchegov:2000hz}.

Let us assume we have two nuclei moving along the light cone,
corresponding to a current
\begin{equation}\label{eq:current}
J^\mu=\delta^{\mu+}\delta(x^-)\rho_{(1)}(\xt)+
\delta^{\mu-}\delta(x^+)\rho_{(2)}(\xt).
\end{equation}
The two colour charge densities $\rho_{(m)}(\xt)$ are, independently for
the two nuclei, drawn from a random
ensemble, which in the original McLerran-Venugopalan model is taken to
be Gaussian:
\begin{equation}\label{eq:rhorho}
\langle \rho^a_{(m)}(\xt)\rho^b_{(m)}(\yt)\rangle=
g^2\mu_{(m)}^2\delta^{ab}\delta^2(\xt-\yt),
\quad m=1,2,
\end{equation}
where $\mu$ is a parameter describing the transverse density of color
charges\footnote{The relation to the convention introduced
in \cite{Krasnitz:2001qu} is $g^2 \mu = \ls$.}.

We then want to find the color fields generated by this current using the
classical equations of motion
\begin{equation}\label{eq:eom}
[D_\mu,F^{\mu\nu}]=J^\nu.
\end{equation}
In the light cone gauge ($A^+=0$ for nucleus (1), $A^-=0$ for nucleus (2))
one first calculates the pure gauge fields corresponding two the two nuclei:
\begin{equation}
A^i_{(m)}(\xt)=\frac{i}{g}U_{(m)}(\xt)\partial_i U_{(m)}^\dag(\xt),\quad m=1,2.
\end{equation}
These depend on the Wilson lines in the covariant gauge,
\begin{eqnarray}\label{eq:wlines}
U_{(1)}(\xt) &=& P \exp \left\{i \int
\ud x^- A^+_{\mathrm{cov}}(x^-,\xt) \right\}
\\ \nonumber
U_{(2)}(\xt) &=& P \exp \left\{i \int
\ud x^+ A^-_{\mathrm{cov}}(x^+,\xt) \right\},
\end{eqnarray}
which can, for infinitely Lorentz-contracted nuclei, be calculated as
\begin{equation}\label{eq:pureg}
U_{(m)}(\xt)=\exp \left\{-ig \frac{\rho_{(m)}}{\nabt^2}(\xt)\right\}.
\end{equation}
In a temporal gauge $A_\tau=0$ the initial condition at $\tau=0$
for the color fields  $\At(\tau,\xt)$ and $A_\eta(\tau,\xt)$ is given by these
pure gauge fields corresponding to the two nuclei:
\begin{eqnarray}
A^i(0,\xt)&=&A^i_{(1)}(\xt)+A^i_{(2)}(\xt),\nonumber\\
A^\eta(0,\xt)&=&\frac{ig}{2}[A^i_{(1)}(\xt),A^i_{(2)}(\xt)].
\end{eqnarray}
To find the gauge field in the future light cone $\tau>0$ one then solves
the gauge field equations of motion using these initial
conditions. In the gauge $A_\tau=0$
it is easy to find the Hamiltonian and thus the energy of a
given field configuration. Additionally, fixing the Coulomb gauge
in the transverse plane, $\nabt \cdot \At = 0,$ one can also
define a gluon multiplicity corresponding to the classical fields.
The method of solving the classical Yang-Mills equations numerically
has been developed in \cite{Krasnitz:1998ns}. First numerical results for SU(2)
were found in \cite{Krasnitz:1999wc,Krasnitz:2000gz}
and for SU(3) in \cite{Lappi:2003bi,Krasnitz:2001qu,Krasnitz:2002mn}.\footnote{
See also the erratum \cite{Krasnitz:2003jw}.}
The numerical code and notations in this paper are those used in \cite{Lappi:2003bi}.

\section{Saturation and rapidity dependence}\label{sec:sat}

The saturation scale $\qs$ is an important concept in small~$x$ physics
(see e.g. \cite{Iancu:2002xk,Iancu:2003xm} for a review).
It has been found to depend
on the value of $x$ probed in the process as $\qs^2 \sim x^{-\lambda}$
both in analytical studies of the quantum evolution of the color
sources in the classical field model
\cite{Jalilian-Marian:1997dw,Weigert:2000gi,Iancu:2001ad,Mueller:2001uk,
Iancu:2002tr,Blaizot:2002xy,Rummukainen:2003ns}\footnote{See e.g.
\cite{Iancu:2003xm} for a more comprehensive list of references.}
and in phenomenological studies
of DIS data \cite{Golec-Biernat:1998js,Golec-Biernat:1999qd}.
The value of $\lambda$ found in \cite{Golec-Biernat:1998js}
is $\lambda= 0.277 \dots 0.288$.
In \cite{Kharzeev:2001gp,Hirano:2004rs} the values used are
$\lambda= 0.25 \dots 0.3$. In presenting the numerical results in
\se\ref{sec:results} we shall keep $\lambda$ arbitrary and use it only
in \se\ref{sec:pheno} when comparing with experimental results and other
theoretical calculations.

In the McLerran-Venugopalan model
the saturation scale $\qs$ is related to the strength of the classical sources by
\begin{equation}\label{eq:lsqs}
\qs^2 = \frac{g^4 \mu^2 \ca }{4 \pi}
\ln \left( \frac{g^4 \mu^2}{\lqcd^2} \right).
\end{equation}
Note that this is the saturation scale as defined in e.g. \cite{Kovchegov:2000hz},
and differs from the one used in the context of DIS (e.g.
\cite{Rummukainen:2003ns})
by the color factor $\ca$ vs. $\cf$\footnote{
This difference is discussed in more detail in
e.g. \cite{Iancu:2003xm,Mueller:2001fv}. }.
The relation \nr{eq:lsqs} can be derived by calculating the correlator
of the Wilson lines, \eq\nr{eq:wlines}, in the adjoint representation
(see \cite{Jalilian-Marian:1997xn} for the detailed calculation).
Up to a logarithmic uncertainty which we will neglect in this study,
the strengths of the classical sources, $g^2 \mu_{(1,2)},$ should
thus also depend on the $x$ probed in nuclei (1) and (2) respectively.
In the perturbative weak field limit of the model we are considering
here the partons are produced  in a \mbox{2 $\to$ 1} process, with
a gluon produced at a given $\ptt$ and $y$ coming from gluons in the
initial nuclear wave functions at \cite{Gribov:1984tu}:
\begin{equation}\label{eq:x12}
x_{1,2} = \ptt e^{\pm y}/\roots.
\end{equation}
Assuming that the dominant transverse momentum scale
$\langle \ptt \rangle$ does not depend on $y$ one is lead to assume that
to calculate gluon production at a rapidity $y$ one must take
\begin{equation}\label{eq:mulambda}
\mu^2_{(1)} = \mu^2 e^{\lambda y}\textrm{ and }
\mu^2_{(2)} = \mu^2 e^{-\lambda y},
\end{equation}
where $\mu_{(1,2)}$ are the source strengths appearing in the correlator
in \eq\nr{eq:rhorho}.
Note that the geometric mean
$\mu \equiv \sqrt{\rule{0pt}{1.4ex}\mu_{(1)} \mu_{(2)} }$
is the quantity appearing in \eqs\nr{eq:deffe}, \nr{eq:deffn}.
Our assumption that  $\langle \ptt \rangle$ does
not depend on $y$ is valid only at small $y$ in AA-collisions
where, by symmetry, one expects
$\langle \ptt \rangle_y \sim
\left( 1 + \mathcal{O}(\lambda^2 y^2) \right)
\langle \ptt \rangle_{y=0} .$ Further away
from central rapidities one could assume $\langle \ptt \rangle$
to depend on one of the saturation scales of the two nuclei,
and the relation \nr{eq:mulambda} cannot be used any more.
A separate question from the selfconsistency of the calculation discussed
above is whether $\roots$ at RHIC is high enough
for this model to be applicable; we will return to this question in
\se\ref{sec:pheno}.

\section{Results}\label{sec:results}

We calculate the transverse energy and multiplicity per unit rapidity as in
\cite{Lappi:2003bi}, and show our results in terms of
the dimensionless ratios
\begin{equation}\label{eq:deffe}
f_E = \frac{1}{g^4 \mu^3 \pi \ra^2} \frac{\ud E}{\ud \eta}
\end{equation}
and
\begin{equation}\label{eq:deffn}
f_N = \frac{1}{g^2 \mu^2 \pi \ra^2} \frac{\ud N}{\ud \eta}.
\end{equation}
These depend on $\lambda y$ through the source strength (by \eq\nr{eq:mulambda})
and the dimensionless parameter characterising the field strength
$g^4 \mu^2 \pi \ra^2$. We shall consider $g=2$ and $\pi \ra^2 = 140 \fm^2$ as
constants and vary the source strength by using different values of $\mu.$
As in \cite{Lappi:2003bi}, the numerical computation presented here is
done assuming
cubic nuclei, whose projection to the transverse plane fills the whole
2-dimensional lattice (of transverse area $\pi \ra ^2$),
enabling the use of periodic boundary conditions.

\begin{figure*}
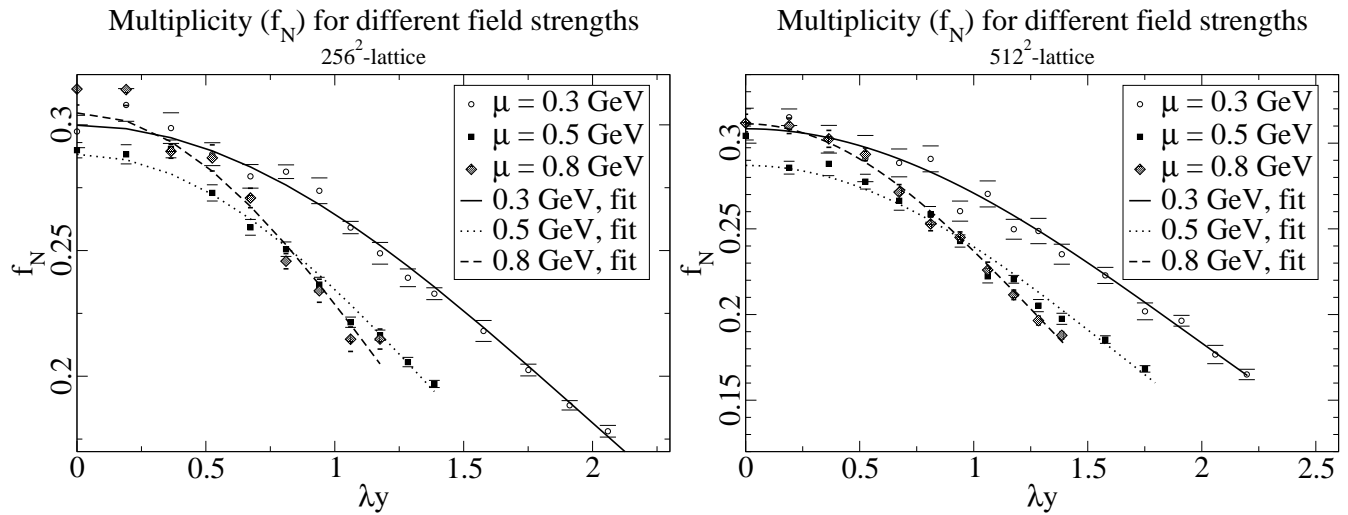

\includegraphics[width=0.49\textwidth]{n256.eps}
\includegraphics[width=0.49\textwidth]{n512.eps}
\caption{\label{fig:n}
The dimensionless ratio $f_N$ for different rapidities
on two different lattice sizes (left: $256^2$, right: $512^2$)
and for different field strengths. Also shown are
the Gaussian fits with the widths in Table \protect\ref{tab:sigma}.}
\end{figure*}

\begin{figure*}
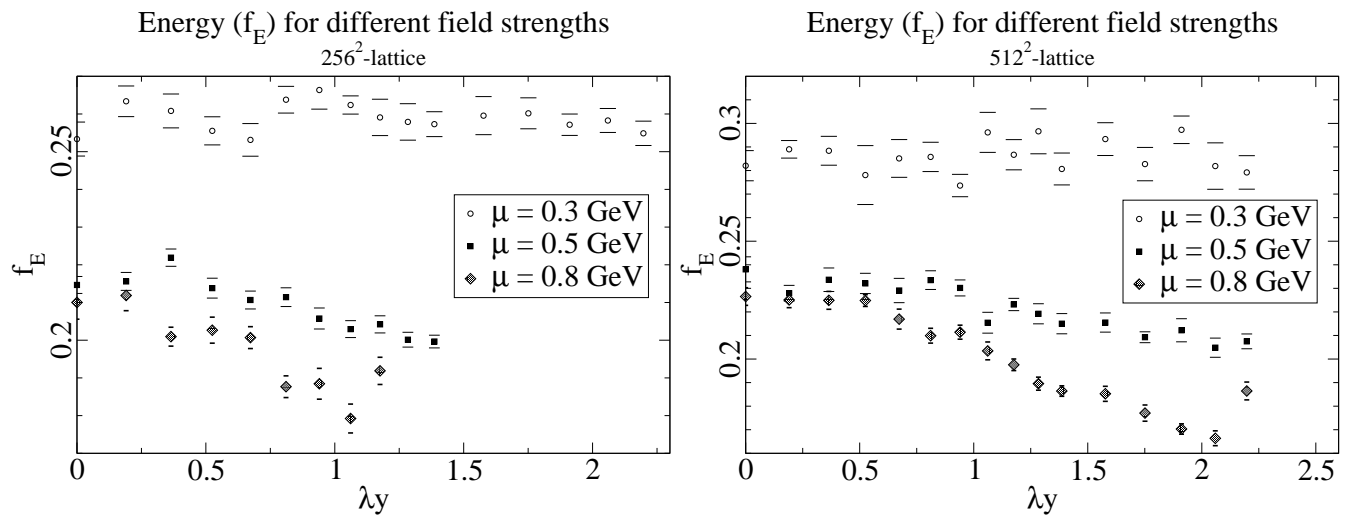

\includegraphics[width=0.49\textwidth]{en256.eps}
\includegraphics[width=0.49\textwidth]{en512.eps}
\caption{\label{fig:en}
The dimensionless ratio $f_E$
on two different lattice sizes (left: $256^2$, right: $512^2$)
and for different field strengths.}
\end{figure*}

\begin{figure}[!htb]
\begin{center}
\includegraphics[width=0.45\textwidth]{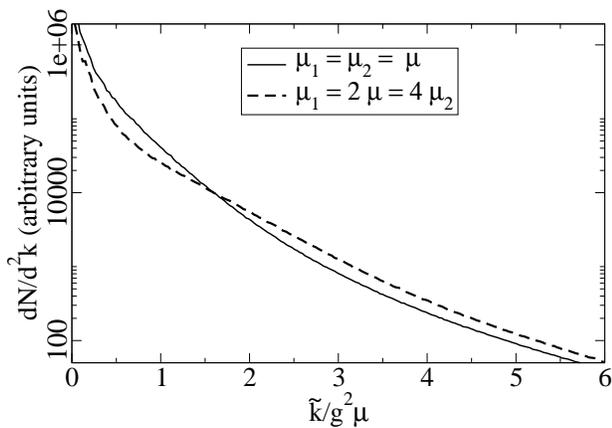}
\end{center}
\caption{\label{fig:spect}
The spectra $\ud N / \ud ^2 \kt$ of produced gluons for a symmetric
collision ($\mu_{(1)} = \mu_{(2)}$) and an asymmetric one
($\mu_{(1)} < \mu_{(2)}$) plotted as a function of $\ktil/g^2 \mu,$ where
$\ktil^2 = 4 \sum_i \sin^2 k_i/2.$
}
\end{figure}

Figure \ref{fig:n} shows the behaviour of $f_N$ and \fig\ref{fig:en} of
$f_E$ as a function of $\lambda y$ for transverse lattices of $256^2$ and
$512^2$ points. One can see that whereas the multiplicity decreases
slowly
with rapidity, the transverse energy does not change significantly,
meaning that the energy per particle increases. This increase is so slow,
however, that our approximation in \se\ref{sec:sat}
of considering $\langle \ptt \rangle$ roughly independent of $y$ is justified
for $\lambda y \lesssim 1$. The behavior of the
gluon spectrum is illustrated in \fig\ref{fig:spect}, which shows the
$\kt$-distibution of gluons for two different rapidities.
This hardening of the spectum is an inherent feature in a
distribution depending on two transverse momentum scales $g^2 \mu_{(1,2)}$.
Note that the presence of two widely different momentum scales also
poses difficulties for the numerical calculation, as the harder scale
approaches the lattice ultraviolet cutoff and the softer scale
approaches the infrared divergent weak field result. For this reason,
although one can use this method to study gluon spectra and the Cronin
effect in ``pA''-collisions (with one of the sources very weak), as
has been done in \cite{Krasnitz:2002mn}, it is perhaps not realistic
to calculate integrated multiplicities for the ``pA''-case.
Instead, saturation physics in
``pA''-collisions\footnote{Actually the relevant RHIC experiments, see e.g.
\cite{Back:2003hx,Arsene:2004ux,Adler:2003ii,Adams:2003im}, are d+Au-collisions.}
can be studied using
a qualitatively different treatment for the proton  and the nucleus (see e.g.
\cite{Kharzeev:2002pc,Kharzeev:2003wz,Baier:2003hr,Blaizot:2004wu}).

We fit the rapidity dependence of the multiplicity with a Gaussian form
\begin{equation}\label{eq:gauss}
f_N = f_0 \exp\left(-\frac{y^2}{2 \sigma^2}\right)
= f_0 \exp\left(-\frac{(\lambda y)^2}{2 (\lambda \sigma)^2}\right).
\end{equation}
Because the result only depends on the combination $\lambda y$,
with $\lambda$ arbitrary so far, the result from the fit is actually
the combination $\lambda \sigma $.
The values of $\lambda \sigma$ from the fits
are listed in Table \ref{tab:sigma}. One sees that there is a small dependence
lattice size and a surprisingly strong dependence on the field strength
$g^4 \mu^2 \pi \ra^2$. The width $\lambda \sigma$ seems to depend on the
field strength more than $f_E(y=0)$ and $f_N(y=0)$.
This means that
 $\lambda \sigma $ is more sensitive to the lattice infrared cutoff than
the energy and multiplicity themselves.
Note that at least in this context there is no fundamental reason for the
Gaussian form, \eq\nr{eq:gauss};
 by symmetry the distribution must be even in $y$ and here we are mainly
interested in the second derivative at $y=0.$ Because of the numerical reasons
discussed above we do not want to go to very large values of $y$ and
a form like like $[1+y^2/(2n\sigma^2)]^{-n}$ for some $n$ would
be just as good. We choose the Gaussian because it has also been used in
experimental \cite{Bearden:2004yx} and other theoretical studies
\cite{Eskola:2002qz}. The rapidity distribution of the energy
does not seem to naturally lend itself to a simple fit that would illustrate
its structure any more than \fig\ref{fig:en}.

\begin{table}
\begin{center}
\begin{tabular}{|c|r||c|c|}
\hline
$\mu$ & $g^4 \mu^2 \pi \ra^2 $
& $\lambda \sigma$, $256^2$-lattice & $\lambda \sigma$, $512^2$-lattice \\
\hline \hline
0.3 GeV & 5184 & 1.99 $\pm$ 0.02 &  1.96 $\pm$ 0.03 \\
\hline
0.5 GeV & 14400 & 1.56 $\pm$ 0.02 & 1.66 $\pm$ 0.02  \\
\hline
0.8 GeV &  36864 & 1.32 $\pm$ 0.03 & 1.35 $\pm$ 0.02 \\
\hline
\end{tabular}
\end{center}
\caption{The widths of the Gaussians $\lambda \sigma$
fitted to the multiplicities in \fig\protect\ref{fig:n}.
}\label{tab:sigma}
\end{table}

\section{Comparison with experimental results and other theoretical
calculations}\label{sec:pheno}
It was found in \cite{Lappi:2003bi} that assuming ideal hydrodynamical expansion
and consequently entropy conservation RHIC multiplicities are best reproduced
by taking $g^2 \mu \approx 2 \gev$ i.e. $\mu \approx 0.5 \gev$.
Taking the corresponding value
$\lambda \sigma \approx 1.66 $ from Table \ref{tab:sigma} and
$\lambda \approx 0.25 \dots 0.3$ one gets $\sigma \approx 5.5 \dots 6.6.$

The dependence of the saturation scale on rapidity has been exploited to
study heavy ion phenomenology also by Kharzeev \& Levin \cite{Kharzeev:2001gp}.
The authors use a saturation-inspired gluon distribution
and perform a pertubative calculation of gluon production.
The same approach has also been used by Hirano \& Nara~\cite{Hirano:2004rs}
to obtain initial
conditions for a hydrodynamical calculation. Estimating from \fig1a
of \cite{Hirano:2004rs} the width of the distribution seems to be
$\sigma \approx 3$, but the shape seems to be flatter
than a Gaussian of this width at small $y$ and
fall more rapidly at large $y$ than a Gaussian.

Our calculation differs from these papers in that we are performing a
numerical calculation by soving the classical field equations, not a
perturbative calculation using unintegrated gluon distributions.
One advantage in our calculation is that the multiplicity comes out
naturally as a smooth function in $y$ around $y=0$, without the
$e^{-\lambda |y|}$-discontinuity in the first derivative in
\cite{Kharzeev:2001gp}. In a perturbative
calculation one can try to incorporate features of high-$x$ physics, such
as the $(1-x)^4$-behaviour of the parton distributions imposed by hand in
\cite{Kharzeev:2001gp,Hirano:2004rs}, whereas our calculation stays
within the framework of a small $x$ model without including this kind
of effects. This
means that our calculation is limited to regions around $y=0,$ where
the fields of both nuclei are strong and can be treated classically,
and such values of $\roots$ that large $x$ effects are not important
at midrapidity.

These results can also be compared to the pQCD+saturation model calculation
\cite{Eskola:2002qz} result of $\sigma = 5.9.$ Whereas in
\cite{Eskola:2002qz} $\sigma$ increases slightly when the saturation scale
grows, in our calculation it decreases and the distribution
becomes more peaked.

\begin{figure*}
\includegraphics[width=0.98\textwidth]{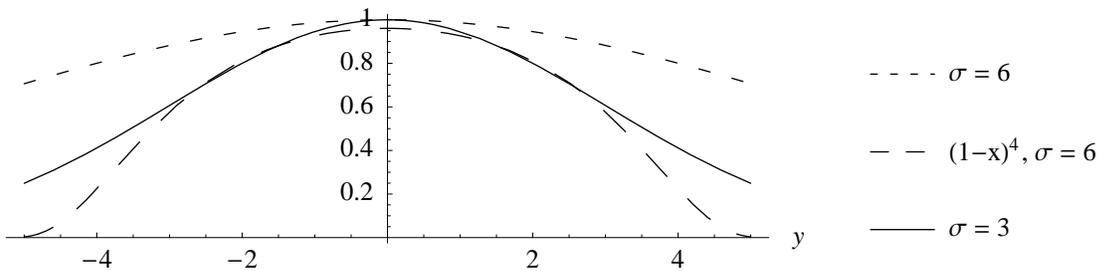}
\caption{\label{fig:oneminusx}
The first dashed curve labeled ``$\sigma=6$'' is the Gaussian
$e^{-y^2/(2 \cdot 6^2)}$. The second curve, labeled ``$(1-x)^4, \quad \sigma=6$''
is the same Gaussian multiplied by $(1-e^y/200)^4 (1-e^{-y}/200)^4$,
where the factor 1/200 comes from assuming $\roots/A = 200 \gev$
and $\langle \ptt \rangle \approx 1 \gev$. The solid curve is
$e^{-y^2/(2 \cdot 3^2)}$, a Gaussian with width $\sigma=3.$
This figure is not intended as a reproduction of the calculation of
\protect\cite{Kharzeev:2001gp,Hirano:2004rs},
but rather as a qualitative demonstration.
}
\end{figure*}

It is not straightforward to compare this calculated initial
state gluon distribution to measured rapidity distributions in the final
state. The approach of \cite{Kharzeev:2001gp} is to assume that they are
equal and only correct for the transformation between rapidity and
pseudorapidity; in the hydrodynamical calculation of
\cite{Eskola:1997hz} the distribution
broadens, but only slightly, during the hydrodynamical evolution. The result of
the BRAHMS collaboration~\cite{Bearden:2004yx}
for the rapidity distribution of charged pions
is $\sigma \approx 2.3,$ which is close to the result of
\cite{Hirano:2004rs} and considerably less than our result.
This could be interpreted as an indication that the main features
of the rapidity dependence of particle production in AA-collisions at
RHIC are not dominated by saturation physics but by the large $x$ behavior
of the parton distribution functions that cause the multiplicity to fall
faster for large rapidities than the classical field model would suggest.
The crucial importance of the factor $(1-x)^4$ in the distribution functions
of \cite{Kharzeev:2001gp,Hirano:2004rs} is illustrated in
\fig\ref{fig:oneminusx}, which shows how, at RHIC energies,
this factor can transform a broad Gaussian with $\sigma = 6$ into a
form that is closer to $\sigma=3$.
Note that the situation at the LHC  will be quite
different as the values of $x$ probed at central rapidities
will be an order of magnitude smaller than at RHIC and thus much less
influenced by large $x$ physics. Another way of viewing this
is that energy conservation limits the size of a rapidity
plateau at central rapidities and forces the multiplicity to decrease
faster with increasing $|y|$ \cite{Eskola:2002qz}.

In this calculation the average transverse momentum of the gluons,
$\langle \ptt \rangle = \frac{f_E}{f_N}g^2\mu$ increases for larger rapidities.
The experimentally measured mean $\ptt$ in the final state is
approximately constant or decreases.
Although one can, in the hydrodynamic scenario,
argue that a large part of the transverse energy goes
into the longitudinal expansion of the system, it is hard to understand
how the energy could decrease by a larger amount for larger rapidities.
In the calculation of \cite{Hirano:2004rs} the transverse
energy decreases faster for larger $y$ due to the explicit
$(1-x)^4$-factor and momentum cutoffs in the transverse integration.

\section{Conclusion and outlook}\label{sec:conclusion}

We have presented a numerical calculation of the rapidity dependence of
gluon production in heavy ion collisions. The calculation
is performed in a $2+1$-dimensional classical Yang-Mills model
assuming that the strength of the classical sources vary with rapidity
according to the simple relation $g^2 \mu \sim e^{\pm \lambda y}$.
Our result is that the multiplicity is approximately a very broad
Gaussian in rapidity with a width $\sigma \approx 6$.
The distribution resulting from the classical field model
is broader than the one observed experimentally. This could indicate
that the rapidity distribution of particle production in heavy ion collisions
at RHIC experiments depends more on the $(1-x)^4$-like
large $x$ behavior of parton distributions than actual saturation physics,
i.e. the $\qs^2 \sim x^{-\lambda}$-dependence of the saturation scale.
At the LHC the situation could be quite different.
Note, however, that our calculation only applies to central rapidities in
AA-collisions and does not address pA-collisions.

The transverse energy produced
in this model does not  decrease for larger rapidities as fast
as one would physically expect. This can be due to the model
not including some relevant large $x$ physics.
Another potential reason could be that
a 2+1-dimensional model is not sufficient to
capture all the aspects of the longitudinal dependence of the problem,
as is the case e.g. for quark pair production \cite{Gelis:2004jp}.
This will be understood better if one is able to perform a full
3+1-dimensional classical field computation with a more detailed
inclusion of the quantum evolution of the sources. This calculation
would also be important for understanding the subsequent thermalisation
of the gluons produced in the collision.

\begin{acknowledgments}
The author is thankful to K. Kajantie for advice and a careful reading of the
manuscript and to K. Tuominen for comments.
This work was supported by the Magnus Ehrnrooth Foundation,
the Finnish Cultural Foundation and the Academy of Finland,
Contract n:o 77744.
\end{acknowledgments}

\bibliographystyle{h-physrev4}
\bibliography{spires}

\end{document}